\documentclass{article}

\usepackage{arxiv}

\usepackage[utf8]{inputenc} 
\usepackage[T1]{fontenc}    
\usepackage{hyperref}       
\usepackage{url}            
\usepackage{booktabs}       
\usepackage{amsfonts}       
\usepackage{nicefrac}       
\usepackage{microtype}      
\usepackage{calc}
\usepackage{epsfig}
\usepackage{subcaption}
\usepackage{calc}
\usepackage{amssymb}
\usepackage{amstext}
\usepackage{amsmath}
\usepackage{amsthm}
\usepackage{multicol}
\usepackage{pslatex}
\usepackage{apalike,pifont}
\usepackage{graphicx}
\usepackage{doi}
\usepackage{cite}
\usepackage{amsfonts}
\usepackage{algorithmic}
\usepackage{graphicx}
\usepackage{textcomp}
\usepackage{multirow}
\usepackage{xcolor}
\usepackage{booktabs}
\usepackage{hyperref}
\usepackage{lscape}

\title{Conceptualising an Anti-Digital Forensics Kill Chain for Smart Homes}

\author{ \href{https://orcid.org/0000-0002-7045-0213}{\includegraphics[scale=0.06]{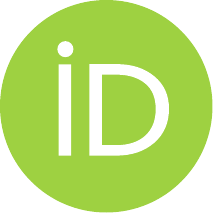}\hspace{1mm}Mario Raciti} \\
	IMT School for Advanced Studies Lucca\\
	Lucca, Italy \\
        Dipartimento di Matematica e Informatica\\
	Università di Catania\\
	Catania, Italy \\
	\texttt{mario.raciti@imtlucca.it} \\
}

\renewcommand{\shorttitle}{Conceptualising an Anti-Digital Forensics Kill Chain for Smart Homes}

\begin{document}
\maketitle

\begin{abstract}
The widespread integration of Internet of Things (IoT) devices in households generates extensive digital footprints, notably within Smart Home ecosystems. These IoT devices, brimming with data about residents, inadvertently offer insights into human activities, potentially embodying even criminal acts, such as a murder. As technology advances, so does the concern for criminals seeking to exploit various techniques to conceal evidence and evade investigations.
This paper delineates the application of Anti-Digital Forensics (ADF) in Smart Home scenarios and recognises its potential to disrupt (digital) investigations. It does so by elucidating the current challenges and gaps and by arguing, in response, the conceptualisation of an ADF Kill Chain tailored to Smart Home ecosystems. While seemingly arming criminals, the Kill Chain will allow a better understanding of the distinctive peculiarities of Anti-Digital Forensics in Smart Home scenario. This understanding is essential for fortifying the Digital Forensics process and, in turn, developing robust countermeasures against malicious activities.
\end{abstract}


\section{Introduction}
\label{sec:introduction}
Locard's exchange principle holds that \textit{``the perpetrator of a crime will bring something into the crime scene and leave with something from it, and that both can be used as forensic evidence.''}~\cite{locard}. This principle also applies in crimes that involve a digital footprint. From a cybersecurity perspective, Digital Forensics (DF) can be considered as a posteriori type of security measure, i.e., it is triggered after a crime has been committed.

A peculiar and recurrent context in which crimes happen is represented by homes. Nowadays, most domestic environments have several Internet of Things (IoT) devices ranging from voice assistants to smart thermostats and security cameras, whose widespread adoption presents unique challenges from a forensic standpoint.
These devices have become an integral part of many households, offering convenience, connectivity, and features so far unthinkable. Such a digital upgrade has changed house buildings into what is defined as Smart Home, with a consequent growth of opportunities for cybercrimes. In fact, a classical example of cybercrime committed in a Smart Home is given from the violation of a device, leading the attacker to take control of it or steal private data, analysing the case from a privacy perspective. In addition, criminals may also leverage the IoT devices within a Smart Home to conduct physical crimes, e.g., by disabling surveillance cameras or unlocking doors.

Moreover, considering the ambivalence of their nature, ``smart'' devices also provide an opportunity for malicious actors to exploit, manipulate, or erase the digital traces that they store, i.e., audio recordings, video footage, and device interaction logs, thus obstructing the forensic process. These activities fall under the umbrella of the unorthodox discipline of Anti-Digital Forensics (ADF).

The objectives and implications of ADF have become subjects of significant debate within the research community and among experts. A prevailing opinion tends to characterise ADF tools as inherently malevolent in both intent and design, yet an alternative perspective posits them towards a more constructive goal. In fact, this viewpoint, argued by J. Foster and V. Liu at the 2005 BlackHat USA Conference~\cite{blackhat2005}, advocates for the use of ADF tools as a means to highlight deficiencies in Digital Forensics procedures and tools. According to the authors, the exposure to Anti-Digital Forensics prompts increased scrutiny by investigators, thus aiming to improve evidence reliability, enhance forensic tools, and elevate the overall quality of forensic education.

This paper explores the challenges and gaps on the application of Anti-Digital Forensics in Smart Home ecosystems and anticipates the conceptualisation of an ADF Kill Chain for Smart Homes as a response.

\section{Background}
\label{sec:related-work}
The traditional definition of Digital Forensics is summarised by NIST as \textit{``the application of computer science and investigative procedures involving the examination of digital evidence -- following proper search authority, chain of custody, validation with mathematics, use of validated tools, repeatability, reporting, and possibly expert testimony.''}~\cite{nist}. From the definition, it follows the important role of Digital Evidence, which is defined as \textit{``Any probative information stored or transmitted in digital form that a party to a court case may use at trial.''}~\cite{caseyh}. 
Furthermore, Horsman and Sunde~\cite{horseman} positioned Digital Forensics as a subset of forensics science, while Abulaish and Haldar~\cite{abulaish} defined it as ``a systematic process of uncovering a crime through the investigation of digital devices''.

\subsection{Smart Home Forensics}
\label{subsec:smart-home}
The study of IoT Security has seen Sklavos et al.~\cite{8049831} discussing trust, security, and privacy, emphasising the educational value and utility of the concepts and protocols. At the same time, IoT Forensics has gained significant attention in recent years due to the growing prevalence of IoT devices and the increasing sophistication of cybercrimes. While several surveys have explored the field of Digital Forensics in the context of IoT~\cite{10141730}, they often focus on specific aspects, such as Network Forensics, Malware Forensics, or Memory Forensics~\cite{branches}. As a result, there is a need for a more comprehensive overview that covers the forensics topics as a whole in the IoT domain and, in particular, in Smart Home ecosystems. 

\begin{figure}[ht]
    \centering
    \includegraphics[width=0.5\textwidth]{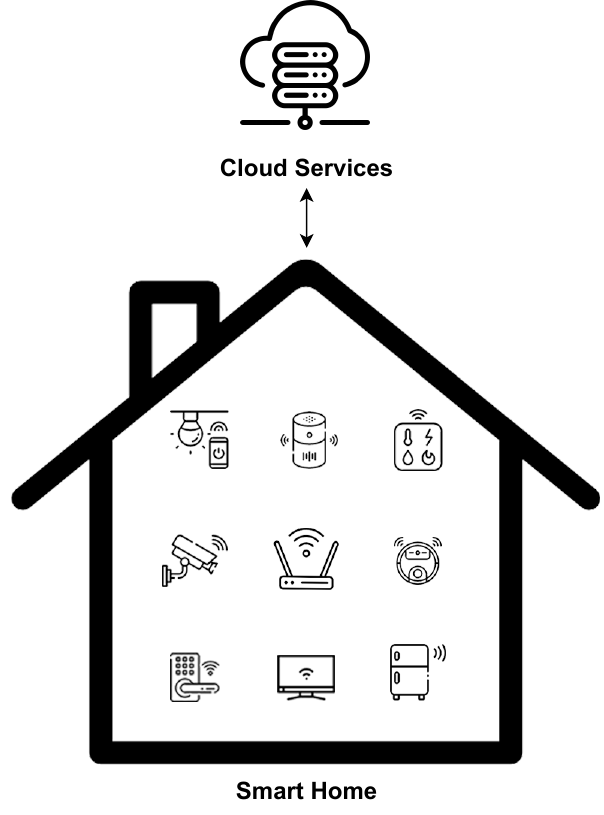}
    \caption{Illustration of a Smart Home Ecosystem.}
    \label{fig:sh}
\end{figure}

A Smart Home ecosystem includes any device that can be monitored or controlled via Internet access. Typically, devices are connected to a central Smart Home hub, i.e., a ``gateway'', and have external interactions with cloud services, e.g., for storage purposes. For the sake of demonstration, Figure~\ref{fig:sh} illustrates a typical Smart Home ecosystem, giving an idea of the complexity and heterogeneity of the elements involved. These include smart doors, smart TVs, smart lights, smart survelliance systems, home assistants, smart sensors, et cetera. A comprehensive understanding of the Smart Home ecosystem serves as a foundational basis, allowing for more targeted and effective studies in various realms, such as Cybersecurity, Privacy and, Digital Forensics.

Ghirardello et al.~\cite{Ghirardello} introduced a reference architecture for smart homes through an exploration of three distinct perspectives within the ecosystem: (i) the functional perspective, encompassing essential operations required for the smart home's regular functioning; (ii) the physical perspective, detailing the physical elements crucial for executing the smart home's functions; and (iii) the communication perspective, outlining the essential protocols for transmitting control and information flows among these components. 

Casey~\cite{casey} emphasised the need for digital investigators to become familiar with smart home systems in order to understand their involvement in criminal activity and the type of information they contain. Furthermore, Kim et al.~\cite{electronics9081215} focused on obtaining, categorising, and examining Smart Home data from notable devices available on the market, i.e., Google Nest Hub, Samsung SmartThings, and Kasa Cam, for forensic purposes. The study scrutinised the collected Smart Home data through companion apps, web interfaces, and APIs to pinpoint significant information applicable to forensic investigations. Also, Awasthi et al.~\cite{AWASTHI2018S38} presented the Almond Smart Home Hub as a potential goldmine for digital forensics, with its centralised data collection. In addition, Lee et al.~\cite{Lee2020} proposed a blockchain-based Smart Home gateway architecture that ensures data integrity and availability, in prevention of data forgery. James~\cite{9108938} developed an intrusion prevention system that can detect and protect against cyber-attacks in Smart Home ecosystems. Anthi et al.~\cite{8753563} introduced a supervised intrusion detection system specifically for Smart Home IoT devices, which can effectively distinguish between benign and malicious network activity. Moreover, Forfot and Østby \cite{10.1007/978-3-030-71711-7_5} suggested a risk assessment model for Digital Forensic Readiness in IoT.

\subsection{Anti-Digital Forensics}
\label{subsec:adf}

Various studies have investigated Anti-Digital Forensics techniques employed by cyber-criminals to hide their activities, but a comprehensive analysis of the various existing Anti-Forensics techniques is often lacking. The discussion surrounding Anti-Forensics (AF) has had a more pronounced impact within law enforcement circles than in the scientific community~\cite{conlan2016anti}. Harris~\cite{harris2006arriving} defined Anti-Forensics as: \textit{``methods used to prevent (or act against) the application of science [...] enforced by police agencies.''}.

\begin{figure*}[ht]
    \centering
    \includegraphics[width=\textwidth]{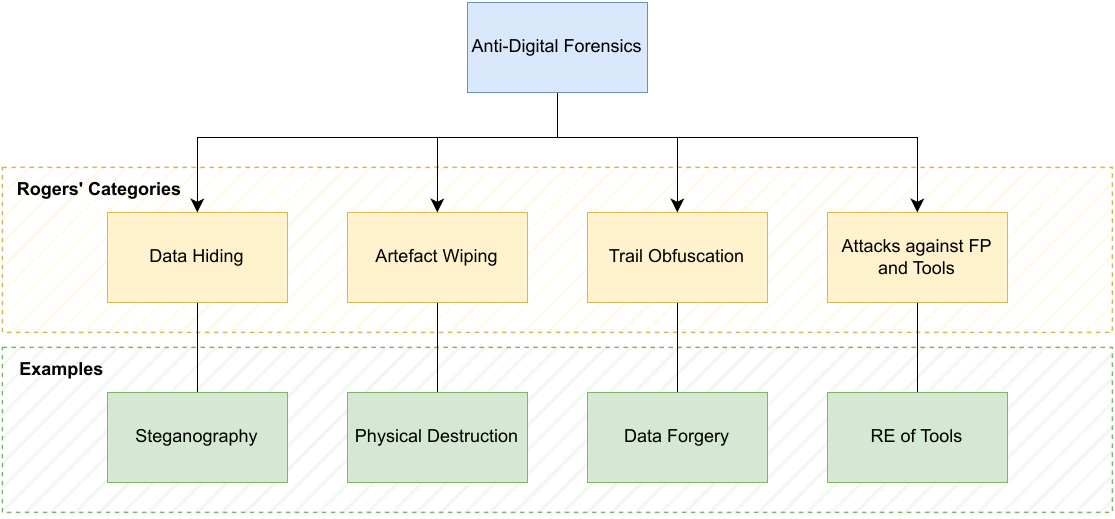}
    \caption{Anti-Digital Forensics Taxonomy and Examples.}
    \label{fig:adf}
\end{figure*}

Literature abounds in numerous definitions of Anti-Digital Forensics, but one of the more widely known and accepted comes from Rogers: \textit{``Attempts to negatively affect the existence, amount and/or quality of evidence from a crime scene or make the analysis and examination of evidence difficult or impossible to conduct.''}~\cite{rogers}. Rogers also proposed a widely adopted taxonomy for the categorisation of ADF techniques: data hiding, artefact wiping, trail obfuscation and attacks against the forensic process and tools. Figure~\ref{fig:adf} depicts Rogers' taxonomy and provides an example for each category: steganography represents a method for data hiding, physical destruction is an extreme practice for artefact wiping, data forgery is a common method for trailing obfuscation, and reverse engineering of forensic tools allows spotting weaknesses and/or vulnerabilities that can lead to hinder the tool effectiveness.

In an Anti-Digital Forensics domain the primary objective is to break the forensic process, thereby such practices can be summarised as the dichotomous counterpart to the ``Forensic Readiness'' (FR) principle, coined by Tan~\cite{SACHOWSKI201645}.
Forensic Readiness was standardised within the Digital Forensic Investigation Readiness Process (DFIRP) model in ISO/IEC 27043:2015 standard. A trending approach to FR, proposed by Rahman et al. \cite{7420536} and conceptually similar to Security-by-design, is Forensic-by-Design (FbD), which aims to integrate forensic requirements into every relevant phase of a system design and development stages, ultimately to obtain ``Forensic-ready'' systems by continuously reviewing the desired state of forensic readiness.

Alenezi et al. \cite{complexis19} advanced a review of challenges and future directions in IoT Forensics, with the inclusion of Anti-Digital Forensics as one of the major challenges. Furthermore, Jean-Paul et al.~\cite{jean} discussed the rise of the Anti-Anti-Forensic protection mechanism against Anti-Forensics activities specifically in IoT systems.

\section{Motivation}
\label{sec:motivation}
As Section~\ref{subsec:smart-home} outlined, there has been some attention directed towards Smart Home Forensics. However, a noticeable disparity arises when considering the scarcity of studies dedicated to Anti-Digital Forensics, especially in relation with Smart Home ecosystems, thereby indicating a discernible gap in the scientific community and among law enforcement and forensics experts.
A critical lack in the current research landscape is given by the absence of guidelines or standardised frameworks that incorporate the steps of Anti-Digital Forensics in Smart Home scenarios. In fact, while the traditional Cyber Kill Chain framework~\cite{lockheed}, developed by Lockheed Martin, is widely used to understand the stages of cyber-attacks and respond effectively, it is not specifically designed for IoT devices, thus failing to address the unique challenges and intricacies associated with Smart Home ecosystems. The same argument holds for the MITRE duality kill chains ATT\&CK~\cite{attack} and D3FEND~\cite{defend}. Moreover, the PEnetration Testing the Internet of Things (PETIoT)~\cite{BELLA2023100707} framework advances a kill chain for IoT devices, but it focuses on cybersecurity aspects, rather than forensics. Hence, the general motivation for this paper to raise the problem and initiate research on the topic.

\section{An Ideal Kill Chain}
\label{sec:kill-chain}
The term \textit{kill chain} is a military concept which identifies the structure of an attack. It typically consists of: identification of target, dispatching of (military) forces to target, initiation of attack on target, destruction of target. Conversely, the idea of ``breaking'' an opponent's kill chain is a method of defence or pre-emptive action. Following our argument, we set the following research questions:

\begin{quote}
RQ \textit{What are the Anti-Digital Forensics steps in a Smart Home ecosystem?}
\end{quote}

To go about such a question, we aim to conceptualise an Anti-Digital Forensics Kill Chain that is tailored for Smart Home ecosystems. The aim of the ADF Kill Chain is twofold. The first is that malicious actors can leverage it as a tool for the exploitation of forensic vulnerabilities. Just as a smoke screen is used in a traditional military action to obscure visibility and confuse the enemy, the Kill Chain supports malicious actors in the creation of a (virtual) veil of confusion within a Smart Home. The secondary aim of the Kill Chain relies on understanding adversary tactics to empower law enforcement to counter those efforts, thereby enhancing forensic investigation effectiveness and ensuring the unimpeded pursuit of truth and justice.

\subsection{Research Goals}
\label{subsec:research-goals}
This paper tentatively outlines the following research goals, acknowledging that they represent an initial stage of inquiry.

\paragraph{Review of ADF in Smart Home.}
Investigate and document existing Anti-Digital Forensics techniques that can be specifically applied to Smart Home ecosystems. Identify the methods malicious actors employ to exploit vulnerabilities within the forensic process, including tampering with evidence, identity concealment, and misleading investigators. Understand the feasibility of these techniques in the context of Smart Home ecosystems.

\paragraph{Intersection of Privacy and ADF.} Examine the current regulatory landscape, with a focus on privacy, and evaluate the differences between legitimate privacy-preserving techniques implemented in Smart Home IoT devices and illicit ADF activities. Identify a common ground where privacy measures can coexist with efforts to counteract malicious ADF techniques.

\paragraph{Integration of AI for ADF in Smart Home.} Investigate the potential role of Artificial Intelligence in support of Anti-Digital Forensics activities in Smart Home ecosystems. Explore opportunities and limitations of existing tools.

\paragraph{Design of ADF Kill Chain for Smart Home.} Develop an Anti-Digital Forensics Kill Chain tailored to Smart Home ecosystems. Identify and categorise the key steps in the proposed ADF Kill Chain, considering the peculiar elements presented by the Smart Home context.
\paragraph{Case studies and real-world applications.} Conduct case studies to illustrate real-world instances of Anti-Digital Forensics in Smart Home ecosystems. Analyse these cases to identify patterns, tactics, and potential variations in ADF techniques employed by adversaries.

\subsection{Expected Challenges}
\label{subsec:expected-challenges}
This paper anticipates the following challenges:

\begin{enumerate}
    \item \textbf{Device heterogeneity}. Smart Home ecosystems encompass various devices with different types of architectures, operating systems, communication protocols, and storage mechanisms. Designing a framework that can handle the heterogeneity of these devices and accommodate their unique characteristics poses a significant challenge.

    \item \textbf{Resource constraints and scalability}. Smart home IoT devices often have limited resources in terms of processing power, memory, and storage. Manipulating or erasing digital evidence within these resource constraints require efficient and scalable techniques. Developing methodologies that optimise resource usage while conducting ADF activities poses a significant challenge. 

    \item \textbf{Forensic readiness}. Ensuring the chain of custody and integrity of digital evidence is crucial in forensic investigations. The Kill Chain needs to address challenges related to hinder the integrity and reliability of evidence, striving to contrast the potential forensic readiness of the Smart Home ecosystem, considering the dynamic nature of its devices.

    \item \textbf{Cloud services}. Smart Home IoT devices often leverage cloud services for data storage and synchronisation. The Kill Chain needs to consider challenges for gaining illicit access to data stored in the cloud, for the subsequent manipulation or erasure of such digital evidence.

\end{enumerate}

\subsection{Development Phases}
The conceptualisation of the ADF Kill Chain for Smart Home ecosystems involves a meticulous workflow. This paper anticipates the main phases for the development of the Kill Chain.

The initial phase is represented by the analysis of a standard Smart Home ecosystem from the perspective of an individual who desires protection against a potential digital forensic investigation. Such analysis entails identifying potential vulnerabilities and weak points in the ecosystem where Anti-Digital Forensics could be employed. The malicious individual might consider aspects such as device communication, data storage, and user interactions as potential areas for obfuscation, manipulation, or destruction of evidence to evade forensic scrutiny. To this extent, a survey on Anti-Digital Forensics techniques provides the appropriate knowledge base to elicit the malicious individual's options.



Furthermore, the ADF Kill Chain needs to be aware of the current status of Forensic Readiness and Forensic-by-Design in Smart Home systems. In fact, these embody measures that the Kill Chain must overcome or, at least, it needs to adapt to.

The validation of the ADF Kill Chain stands as a crucial step in establishing its efficacy and relevance within the context of Smart Home ecosystems. This phase involves rigorous testing and assessment to ensure that each stage of the Kill Chain accurately reflects real-world adversarial activities and their impact on digital forensic investigations.

As we shall see below, the next phase involves the practical application of the ADF Kill Chain to a designed case study that mirrors realistic scenarios in Smart Home environments.

In addition, the evaluation phase encompasses the examination of how well the ADF Kill Chain can be executed by adversaries seeking to manipulate or destroy digital evidence within Smart Home IoT devices. This process helps identify potential weaknesses and areas of improvement in the Kill Chain, allowing for iterative refinement.

Assessments on defensive measures against the Kill Chain conclude the flow and ensure an understanding of its (technical, legal, and ethical) limits.

\subsection{Preliminary Conceptualisation}

We anticipate that certain stages within our conceptualised Kill Chain may exhibit similarities with established state-of-the-art kill chains. This expectation stems from the recognition that adversarial strategies often share fundamental principles across different cyber contexts. For instance, the well-known steps of ``Reconnaissance'' and ``Exfiltration'' can arguably be part of a Kill Chain tailored for Anti-Digital Forensics, as fundamental aspects of Vulnerability Assessment and Penetration Testing. By following this argument, in Smart Home ecosystems the Reconnaissance step can be exemplified by the identification of the devices connected, namely device discovery. 

Moreover, in our current research, we have identified some steps that are peculiar to ADF scenarios and, at the same time, differ from the classical steps of a cybersecurity-oriented kill chain. 
While the design, including the detail and ordering, of these steps have yet to be fully conjectured, an overarching conceptualisation is given below.

\paragraph{Step A --- Tampering with Digital Traces}
\, \\ 
\textit{Objective:} Manipulate or erase digital traces to obstruct forensic investigation.
\\
\textit{Activities:} Tampering with audio recordings, video footage, and device interaction logs. Implementing techniques to make forensic analysis challenging.

\paragraph{Step B --- Concealing Identities}
\, \\
\textit{Objective:} Conceal the identity of malicious actors involved in ADF activities.
\\
\textit{Activities:} Masking IP addresses and digital footprints. Falsifying user identities associated with Smart Home devices.

\paragraph{Step C --- Misleading Investigators}
\, \\
\textit{Objective:} Introduce false information to mislead forensic investigators.
\\
\textit{Activities:} Planting deceptive digital breadcrumbs and manipulating timestamps and metadata.

\paragraph{Step D --- Cloud Data Manipulation}
\, \\
\textit{Objective:} Manipulate data stored in cloud services associated with Smart Home devices.
\\
\textit{Activities:} Getting remote access to cloud services where Smart Home data is stored. Tamper with or delete such data remotely, ensuring techniques to avoid logging. \\

\subsection{A Case Study}
\label{subsec:case-study}
For the application of the ADF Kill Chain, we propose a case study on a Smart Home system that resembles the following hypothetical scenarios. A criminal suspect, namely Mr. X, is under investigation for a serious crime. Whether Mr. X is guilty or not, he is aware of the digital footprints left by Smart Home IoT devices and is concerned about the investigation. Therefore, Mr. X seeks to exploit these devices to create a fabricated digital alibi that misrepresents his presence at the crime scene. To achieve this, he devises a plan to tamper with and destroy digital evidence from the Smart Home IoT devices within his residence. Mr. X believes that by evading digital forensic detection, he can create a false narrative to defend himself against the charges. Mr. X's plan involves leveraging the Kill Chain to carry out his digital alibi fabrication scheme.
Following the outcomes from the above scenario, we may model a second scenario featuring an investigator, namely, Mrs. Y, who wants to understand the crucial steps that Mr. X employed to hinder the investigation in the Smart Home crime scene. Thus, Mrs. Y can leverage the Kill Chain for Digital Forensics purposes. 

\section{Conclusions}
\label{sec:conclusions}
This paper provided an in-depth exploration of the challenges and gaps presented by the forensic perspective of Smart Home ecosystems, specifically addressing the connections with the emerging subject of Anti-Digital Forensics. The paper advised a direction to answer the research question by arguing the conceptualisation of an ADF Kill Chain tailored for Smart Home ecosystems.

The ethical implication of a Kill Chain supporting criminals (at least at a first glance) will be mitigated by the dual product of elucidating and understanding potential adversarial steps, hence providing the knowledge to prepare appropriate countermeasures in support of Anti-Anti-Forensics.

The endeavour to fill the void in training, awareness, and knowledge-sharing within the ADF landscape is essential for ensuring the integrity  and efficacy of the Digital Forensics process applied in Smart Home scenarios.
This paper encouraged future research to enhance the comprehension of Anti-Digital Forensics, both in a general context and in its application in Smart Home ecosystems. This will significantly help to elucidate the boundaries of Digital Forensics, with a consequent mitigation of the so-called ``CSI effect.'', i.e., the phenomenon where people's perception of forensics science is influenced by fictional portrayals in popular media, thus improving criminal trials and court proceedings.

\bibliographystyle{unsrtnat}
\bibliography{references}  






\end{document}